\documentclass[aip,apl,reprint]{revtex4-1}
\usepackage{amsmath}
\usepackage{amsfonts}
\usepackage{amssymb}
\usepackage{graphics}
\usepackage{float}
\usepackage{graphicx}
\usepackage{epstopdf}
\usepackage[compact]{titlesec}
\usepackage{natbib}
\usepackage{threeparttable}
\usepackage[titletoc,title]{appendix}
\usepackage{lipsum}
\usepackage{xcolor}
\begin{document}
\title{Anomalous Lorenz number in massive and tilted Dirac systems}
\author{Parijat Sengupta}
\affiliation{Dept. of Electrical and Computer Engineering, Purdue University, West Lafayette, IN, 47907, USA.}
\affiliation{Dept. of Electrical and Computer Engineering and Material Science Division, Boston University, Boston, MA, 02215, USA.}
\author{Enrico Bellotti}
\affiliation{Dept. of Electrical and Computer Engineering and Material Science Division, Boston University, Boston, MA, 02215, USA.}

\begin{abstract}
We analytically calculate the anomalous transverse electric and thermal currents in massive and tilted Dirac systems, using $ \beta $-borophene as a representative material, and report on conditions under which the corresponding Lorenz number $\left(\mathcal{L}_{an}\right)$ deviates from its classically accepted value $\left(\mathcal{L}_{0}\right)$. The deviations in the high-temperature regime are shown to be an outcome of the quantitative difference in the respective kinetic transport expressions for electric $\left(\sigma\right)$ and thermal $\left(\kappa\right)$ conductivity, and are further weighted through a convolution integral with a non-linearly energy-dependent Berry curvature that naturally arises in a Dirac material. In addition, the tilt and anisotropy of the Dirac system that are amenable to change via external stimulus are found to quantitatively influence $ \mathcal{L}_{an} $. The reported deviations from $\mathcal{L}_{0} $ hold practical utility inasmuch as they allow an independent tuning of $\sigma $ and $ \kappa $, useful in optimizing the output of thermoelectric devices.
\end{abstract}
\maketitle

The Lorenz number links the electrical conductivity $\left(\sigma\right)$ of a material to its thermal counterpart via the Wiedemann-Franz law (WFL).~\cite{thesberg2017lorenz} The WFL, for a temperature $ T $, is given as $ \mathcal{L}_{0} = \kappa_{e}/\sigma T $, where the subscript `\textit{e}' denotes the electronic component of the overall thermal conductivity $\left(\kappa = \kappa_{e} + \kappa_{l}\right)$. The lattice contribution to $ \kappa $ is distinguished by the `\textit{l}' subscript. The Lorenz number $\left(\mathcal{L}_{0}\right)$ is $ 2.44 \times 10^{-8} W\Omega K^{-2} $, and holds constant when the electron gas is highly degenerate and the electronic mean free path is equal for electrical and thermal conductivities.~\cite{mizutani2001introduction} The $ \mathcal{L} $ value can, however, undergo alteration~\cite{kim2015characterization,principi2015violation} in a wide variety of situations; for instance, it drops to $ 1.5 \times 10^{-8} W\Omega K^{-2} $ in non-degenerate semiconductors with inelastic acoustic phonon scattering. A recent study on bulk $ WP_{2} $ single crystals demonstrated a Lorenz number heavily mismatched to $ \mathcal{L}_{0} $, and this was shown to occur driven by a strong electron-electron scattering.~\cite{jaoui2018departure} The utility of the WFL-guided $ \mathcal{L}_{0} $ and its variations thereof, however, lie in the `window' it offers to an independent modulation of $\sigma $ and $ \kappa_{e} $, the two material-specific parameters that govern thermoelectric behavior. A thermoelectric energy converter relies on a low $ \kappa $ and a large $ \sigma $ to increase efficiency. In a classical case, where $ \mathcal{L}_{0} $ is a constant, typically, a rise in $ \sigma $ is accompanied by a concomitant increase in $ \kappa $, thereby impeding the full optimization of thermoelectric behavior. For a meaningful thermoelectric tuning, $ \mathcal{L}_{0} $  must be adjustable to preset requirements, and this has been shown viable through tailored electron transport, for example, via controlled scattering events~\cite{pei2011convergence}. 

\begin{figure}
\includegraphics[scale=0.5]{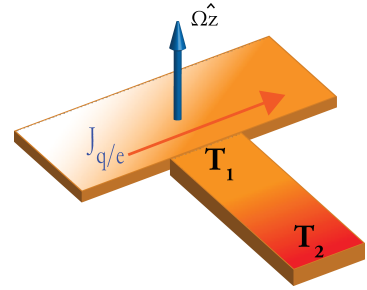}
\vspace{-0.28cm}
\caption{A generalized illustration of the anomalous thermal Hall (ATH) effect induced by the out-of-plane directed and momentum-dependent Berry curvature $ \left(\Omega\left(k\right)\right) $ is shown here. The ATH sets up a transverse temperature difference $\left(T_{2} - T_{1}\right)$ in presence of a longitudinal heat current $\left(J_{q}\right)$. The anomalous Hall conductivity parallels the ATH leading to production of a transverse potential difference when a longitudinal charge current $\left(J_{e}\right)$ flows through the sample.}
\vspace{-0.2cm}
\label{berth}
\end{figure}  

In this letter, we merge such transport techniques usually computed within the framework of Boltzmann equation with wave packet dynamics of Bloch electrons where the topological Berry curvature manifests to quantify the Lorenz number. A key hallmark of such bands is the presence of the momentum-space Berry curvature $\left(\Omega\left(k\right)\right)$ which imparts an `additional' velocity of the form $ \dot k \times \Omega $ to the Bloch electrons, and effectively mirrors the Lorentz force of a magnetic field.~\cite{zhang2016berry} Here, the momentum vector is $ \mathbf{k} $. A finite $ \Omega\left(k\right) $, which can exist in a time reversal symmetry (TRS) broken or inversion asymmetric system has been shown to give rise to anomalous electric and thermal behavior.~\cite{xiao2006berry,onoda2008quantum} The anomalous electric $\left(\sigma_{xy}\right)$ Hall conductivity (\textbf{AHC}) arises in material systems, wherein a non-zero $ \Omega\left(k\right) $ in presence of a longitudinal (\textit{x}-directed) potential gradient induces a transverse (along the \textit{y}-axis) potential difference. The thermal counterpart of this anomalous Hall conductivity (\textbf{THC}) leads to a transverse temperature difference in response to a longitudinal temperature gradient.~\cite{yokoyama2011transverse,zhang2009berry} We seek to map and modulate for a given temperature $ T $, the `anomalous' $ \mathcal{L}_{an} = \kappa_{xy}/\sigma_{xy} T $ for Bloch electrons that live in a topologically non-trivial energy manifold. The linear and tilted bands of borophene around the $ \Gamma $ point of the Brillouin zone conform to such an energy description. The calculations that follow primarily use borophene as the representative Dirac material to unveil the possible `topological' alterations to $ \mathcal{L}_{an} $. 

Briefly, we find that $ \mathcal{L}_{an} $ remains close to $ \mathcal{L}_{0} $ for lower temperatures and is shown to be impacted by the intrinsic tilt and anisotropy of the Dirac dispersion intrinsic to $ \beta $-borophene. The $ \mathcal{L}_{an} $, however, diverges at higher values of temperature. This divergence from $ \mathcal{L}_{0} $ is attributed to the dissimilar character of material-dependent transport expressions appropriate for AHC and THC; these expressions while closely matched at low temperatures, exhibit a large dissimilarity at other thermal regimes. In addition, $ \mathcal{L}_{an} $ goes further adrift of $ \mathcal{L}_{0} $ as a consequence of their weighted integral with the $ \Omega\left(k\right) $ that non-linearly scales with energy $\left(k_{B}T\right)$. The transport component and the $ \Omega\left(k\right) $ operating in tandem in each of AHC and THC further accentuates the overall observed disparity of $ \mathcal{L}_{an} $ vis-\'a-vis $ \mathcal{L}_{0} $ at higher temperatures.

\begin{figure}[t!]
\includegraphics[scale=0.52]{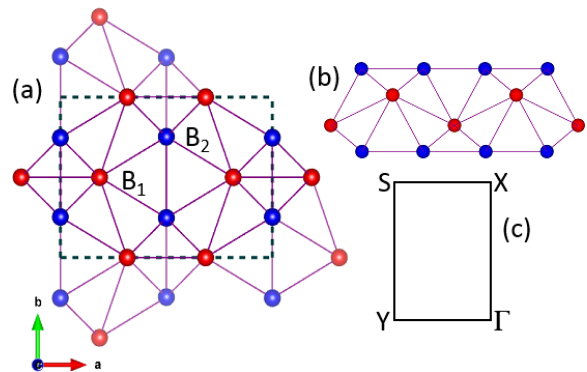}
\caption{The eight-atom simple orthorhombic arrangement in the $ \beta $-borophene unit cell (marked by dotted lines) is illustrated here. There are two inequivalent atom positions indicated by the identifiers, B$_{1}$ and B$_{2}$, in sub-figure (a). The coordinates of B$_{1}$ are $\left(0.185a, b/2, 0.531c\right)$ while those for B$_{2}$ can be written as $\left(a/2, 0.247b,0.584c\right)$; the difference in the \textit{z}-coordinate represents the intrinsic buckling (b). The high-symmetry points in the corresponding Brillouin zone are shown in (c); the tilted Dirac cone in $\beta $-borophene is formed along the $ \Gamma - Y $ direction.}
\label{ubcell}
\end{figure}

The two-dimensional (2D) boron monolayers known as borophenes~\cite{lopez2016electronic,sengupta2018anomalous} have been proposed and synthesized in a variety of allotropes. The electron-deficient boron atom participates in a wide variety of complex bonding patterns from which emerges stable crystal structures such as quasi-planar clusters, cage-like fullerenes, and nanotubes. In particular, a free-standing arrangement of two-dimensional boron atoms with a buckled structure (Fig.~\ref{ubcell}) and an orthorhombic 8-$ Pmmn $ symmetry ($ Pmmn $ represents the space group 59; 8 denotes the number of atoms in the unit cell) was shown to carry anisotropic and tilted Dirac cones. For brevity, we refer to this form as $ \beta $-borophene. We begin by writing the low-energy continuum two-band Hamiltonian~\cite{zabolotskiy2016strain} for $ \beta $-borophene that describes an anisotropic and tilted Dirac crossing along the $ \Gamma $-Y direction in the Brillouin zone (see Fig.~\ref{ubcell}(c)).
\begin{equation}
H = \hbar v_{x}\sigma_{x}k_{x} + \hbar v_{y}\sigma_{y}k_{y} + \hbar v_{t}\sigma_{0}k_{y}.
\label{boeqn}
\end{equation}
In Eq.~\ref{boeqn}, $ \sigma_{x,y} $ are the Pauli matrices denoting the lattice degree of freedom while $ \sigma_{0} $ is  the $ 2 \times 2 $ identity matrix. The direction-dependent velocity terms $\left(\times 10^{6}\,m/s \right)$ as reported in Ref.~\onlinecite{zabolotskiy2016strain} are $ v_{x} = 0.86 $, $ v_{y} = 0.69 $, and $ v_{t} = 0.32 $. Note that anisotropy, $ \nu = v_{x}/v_{y} $, arises since $ \nu \neq 1 $ while $ v_{t} \neq 0 $ ensures a tilt through non-concentric constant energy contours. The dispersion relation using Eq.~\ref{boeqn} is
\begin{equation}
E_{\pm}\left(k\right) = \hbar k\left(v_{t}\sin\left(\theta\right) \pm\sqrt{v_{x}^{2}\cos^{2}\theta + v_{y}^{2}\sin^{2}\theta}\right).
\label{diseq}
\end{equation}
The upper (lower) sign is for the conduction (valence) band in $\beta $-borophene. This basic energy description (Eqs.~\ref{boeqn},~\ref{diseq}) serves as the starting point to examine models of light-matter interaction. 
A diagonalization of the representative borophene Hamiltonian (Eq.~\ref{boeqn}) furnishes a Dirac cone; however, there exist another Dirac cone and the pair are related by symmetry operations; for a group theory analysis of the underlying symmetry, see, for example, Ref.~\onlinecite{zabolotskiy2016strain}. Briefly, the two Dirac cones though identical in dispersion have reversed chirality and marked by exactly opposite tilts. The dispersion (Eq.~\ref{diseq}) is slightly modified by placing a negative sign before $ v_{t}\sin\left(\theta\right) $, the first term within parenthesis.

\begin{figure}[t!]
\includegraphics[scale=0.47]{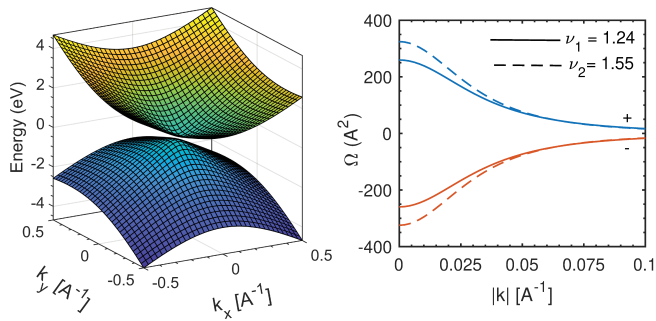}
\caption{The left panel shows the tilted and gapped Dirac cone (DC) for $ \beta $-borophene plotted using Eq.~\ref{diseq}. For visual clarity, we set the gap to an artificial $\Delta = 0.2\, eV $. The anisotropy in the dispersion around the DC is clearly seen. The right panel, assuming an identical $ \Delta $, plots the Berry curvature $\left(\Omega\right)$ of the Bloch electrons that lie close to the DC. We show it for two values of the parameter $ \nu = v_{x}/v_{y} $, which defines the anisotropy of the dispersion. The $ \Omega $ of the occupied bands drives the anomalous thermal and electrical phenomena.}
\label{dberry}
\end{figure}

We noted above that carriers with a finite Berry curvature $\left(\Omega\right)$ lead to anomalous electrical and thermal conductivities. For the Bloch electrons governed by the two-band Hamiltonian in Eq.~\ref{boeqn} to experience a non-zero $ \Omega $, it is necessary to introduce terms that may either break inversion or time reversal symmetry. We introduce a simple inversion symmetry breaking term of the form $ \Delta\sigma_{z} $ to the Hamiltonian. Note that $ \sigma_{z} $ points to the sub-lattice degree of freedom and is therefore time reversal invariant $\left(\sigma_{z}\overset{\mathcal{T}}{\rightarrow} \sigma_{z}\right)$. Here, $\mathcal{T} $ is the time reversal symmetry operator. The term $ \Omega\sigma_{z} $ manifests as a band gap opening in the gapless Dirac cones. For a general two-band model Hamiltonian of the form $ H = \mathbf{\sigma}\cdot \mathbf{d}\left(k\right) + \mathbb{I}\epsilon\left(k\right) $, where $ \mathbf{d} $ is a vector of spin or pseudo-spin, $ \epsilon\left(k\right) $ is a general dispersion term, and $ \mathbb{I} $ is the $ 2 \times 2 $ identity matrix, the Berry curvature for this system is~\cite{fruchart2013introduction} 
\begin{equation}
\Omega_{\mu\nu} = \dfrac{1}{2}\varepsilon_{\alpha\beta\gamma}\hat{d}_{\alpha}\left(\mathbf{k}\right)\partial_{k_{\mu}}\hat{d}_{\beta}\left(\mathbf{k}\right)\partial_{k_{\nu}}\hat{d}_{\gamma}\left(\mathbf{k}\right),
\label{berry}
\end{equation}
where $ \hat{\mathbf{d}}\left(\mathbf{k}\right) = \dfrac{\mathbf{d\left(\mathbf{k}\right)}}{d\left(k\right)} $. Applying this formalism to the $ \beta $-borophene Hamiltonian (Eq.~\ref{boeqn}), and noting that the vector $ \mathbf{d} $ in component notation assumes the form:
\begin{equation}
\mathbf{\hat{d}}\left(k\right) = \dfrac{\hbar}{\sqrt{\left(\hbar v_{x}k_{x}\right)^{2} + \left(\hbar v_{y}k_{y}\right)^{2} + \Delta^{2}}}\left(v_{x}k_{x},v_{y}k_{y}, \Delta\right).
\label{dmer}
\end{equation}
Substituting the $ \mathbf{d} $ vector in Eq.~\ref{berry}, $ \Omega\left(k\right) $ is expressed  as:
\begin{equation}
\Omega\left(k\right) = \mp\dfrac{\hbar^{2}v_{x}v_{y}\Delta}{2\left[\left(\hbar^{2}v_{x}^{2}k_{x}^{2} + \hbar^{2}v_{y}^{2}k_{y}^{2}\right)^2 + \Delta^2\right]^{3/2}}\hat{\mathbf{z}}.
\label{bc}
\end{equation}
The upper (lower) sign is for the conduction (valence) band. The $ \Omega\left(k\right) $ as a momentum-dependent magnetic field points out-of-plane (the \textit{z}-axis) and evidently decays as $ \Delta \rightarrow 0 $. A plot of $ \Omega\left(k\right) $ for $ \vert k \vert $ values centered around the Dirac crossing is shown in Fig.~\ref{dberry}. The $ \Omega\left(k\right) $ as expected peaks around the $ \vert k \vert = 0 $ mark and diminishes as we move farther in momentum-space.

\begin{figure}[t!]
\includegraphics[scale=0.77]{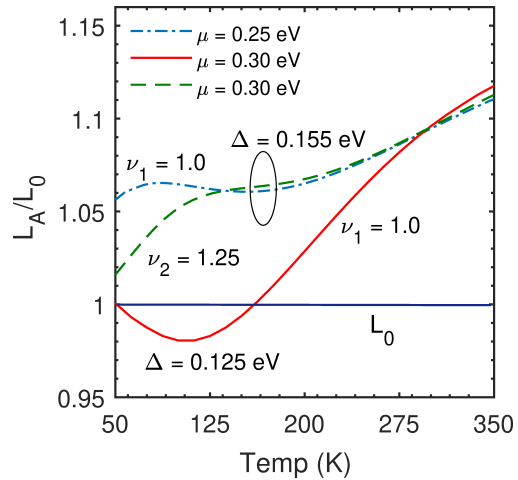}
\vspace{-0.1cm}
\caption{The numerically calculated anomalous Lorenz number $ \left(\mathcal{L}_{an}\right) $ is shown here. For low temperatures, $ \left(\mathcal{L}_{an}\right) $ tracks the classical value $\left(\mathcal{L}_{0}\right)$. The corresponding high temperature values reveal a deviation as discussed in the main text. The $ \mathcal{L}_{an} $ is also computed for two values of $ \nu = v_{x}/v_{y} $, the anisotropy parameter of the the Dirac bands. The anisotropy influence the occupancy of energy states around the Fermi level and play a crucial role in the determining the magnitude of the anomalous $ \sigma_{xy} $ and $ \kappa_{xy} $ from which $ \mathcal{L}_{an} $ is derived.}
\label{lan}
\end{figure}

We have at this point gathered the pieces required for a quantitative evaluation of the anomalous thermal and electrical conductivity coefficients and their ratio which gives us $ \mathcal{L}_{an} $. Briefly, we recall that the transverse flow of both electric and thermal currents arise from the curvature of electrons under a magnetic field; for our case, the magnetic field is the momentum dependent $ \Omega\left(k\right) $. In presence of a temperature gradient, the anomalous thermal Hall conductivity is formally defined via the relation $ j_{q,y} = -\kappa_{xy}\nabla_{x}T $. The heat current along \textit{y}-axis is $ j_{q,y} $, and transverse to a temperature gradient vector aligned to the \textit{x}-axis (Fig.~\ref{berth}. This characteristic coefficient of this transverse heat flow is given as~\cite{bergman2010theory,qin2011energy}
\begin{equation}
\begin{aligned}
\kappa_{xy} &= \dfrac{k_{B}^{2}T}{h}\int\dfrac{\mathbf{d}^{2}k}{4\pi^{2}}\sum_{\tau}\Omega_{\tau}\left(k\right)\biggl[\dfrac{\pi^{2}}{3} + k_{B}T\left(E - \mu\right)f \\
&- ln^{2}\left(1 - f\right) - 2Li_{2}\left(1 - f\right)\biggr].
\label{thber}  
\end{aligned}
\end{equation}
In Eq.~\ref{thber}, $ Li_{n}\left(x\right) = \sum_{m = 1}^{\infty}\dfrac{x^{m}}{m^{n}} $ is the poly-logarithmic function. The symbol $ f $ denotes the equilibrium Fermi distribution. Similarly, in the presence of an external electric field E along the \textit{x}-axis, the anomalous Hall current is along the transverse (\textit{y}-axis) direction, from which the the corresponding conductivity is given as (BZ: Brillouin zone)
\begin{equation}
\begin{aligned}
\sigma_{xy} = \dfrac{e^{2}}{\hbar}\int_{BZ}\dfrac{\mathbf{d}^{2}k}{4\pi^{2}}\Omega\left(k\right)f\left(k\right).
\label{hcber}  
\end{aligned}
\end{equation}
The anomalous Lorenz number ratio of $ \mathcal{L}_{an} = \kappa_{xy}/\sigma_{xy}T $ which can be computed by a direct application of Eqs.~\ref{thber} and ~\ref{hcber} is shown in Fig.~\ref{lan} for a pair of Fermi levels and related band gaps. In addition, the anisotropy $\left(\nu\right) $ quantitatively influences $ \mathcal{L}_{an}$. We begin by first noting that at low temperatures, the $ \mathcal{L}_{an} $ values are close to $ \mathcal{L}_{0} $ (shown by the horizontal straight line in Fig.~\ref{lan}). They exhibit a more discernible departure from $ \mathcal{L}_{0} $ at higher temperatures, and in general, evidently deviate from $ \mathcal{L}_{0} $. 

\begin{figure}[t!]
\includegraphics[scale=0.67]{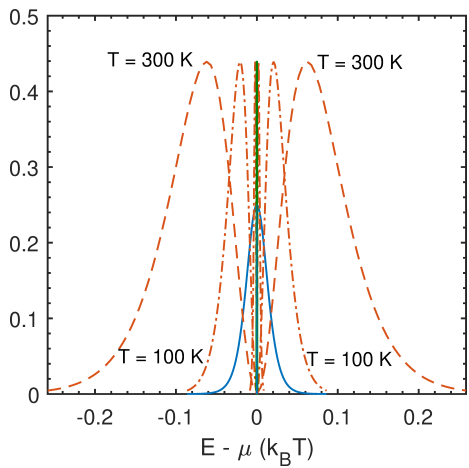}
\caption{The kernel functions, $ \chi_{1} $ (solid curve) and $ \chi_{2} $ (dashed curve), are plotted for $ T = 1, 10, 100, 300\,\mathrm{K} $. At $ T = 1\,K $, $ \chi_{1} $ and $ \chi_{2} $, which are the kinetic transport functions for AHC and THC, respectively, completely overlap and seen as a spike for $ E - \mu \approx 0 $. A double valley structure for $ \chi_{2} $ opens up at $ T = 10\, K $. The spacing between the two valleys is of the order $ k_{B}T $ and for low temperatures are in close proximity. For higher temperatures, the valleys open up and are numerically distinguishable from the single valley of $ \chi_{1} $. At $ T = 100\, K $, the profile of $ \chi_{1} $ is centered around $ \mu $ while the valleys of $ \chi_{2} $ are spread over a larger energy range. The `spread' is typically greater when the temperature rises, as is easily noticeable for the case of $ T = 300\, K $, and contributes to the deviation of $ \mathcal{L}_{an} $ from $ \mathcal{L}_{0} $.}
\label{kerfunc}
\vspace{-0.2cm}
\end{figure}

In the following, we try to explain this anomalous behaviour; to do so, let us begin by rewriting the expression for anomalous thermal Hall conductivity as $ \kappa_{xy} = -\left(k_{B}^{2}T/h\right)c_{2} $, where $ c_{2} $ is~\cite{bergman2010theory}
\begin{equation}
c_{2} = -\int\dfrac{d^{2}k}{4\pi^{2}}\sum_{\tau}\Omega_{\tau}\int_{E-\mu}^{\infty}dE\left(\beta E\right)^{2}\dfrac{\partial f\left(E\right)}{\partial E}.
\label{altdef}
\end{equation}
Note that expanding the expression for $ c_{2} $ in Eq.~\ref{altdef} by inserting the full form for the Fermi distribution gives Eq.~\ref{thber}. We can also analogously write a similar coefficient $ c_{1} $ for the AHC $ \left[\sigma = -\left(e^{2}/h\right)c_{1}\right] $; it is~\cite{bergman2010theory} 
\begin{equation}
c_{1} = -\int\dfrac{d^{2}k}{4\pi^{2}}\sum_{\tau}\Omega_{\tau}\int_{E-\mu}^{\infty}dE \dfrac{\partial f\left(E\right)}{\partial E}.
\label{altdefahc}
\end{equation}
A straightforward comparison of Eqs.~\ref{thber} and ~\ref{altdefahc} shows that the Berry curvature that sums over the occupied bands has two different sets of kernels: $ \chi_{1} = \dfrac{\partial f\left(E\right)}{\partial E} $ and $ \chi_{2} = \left(\beta E\right)^{2}\dfrac{\partial f\left(E\right)}{\partial E}$ for the AHC and THC, respectively.~\cite{behnia2015fundamentals} The kernel for AHC $\left(\chi_{1}\right)$ has a single peak that receives contribution only close to $ \mu $; in contrast, $\chi_{2} $ has double valleys supported by states both below and above $\mu $ with identical sign. This behavior is illustrated in Fig.~\ref{kerfunc}. The solid (blue) curve is the single-peaked plot for the AHC kernel, $ \chi_{1} $, in units of $ 1/k_{B}T $ at $ T = 100\, K $. The corresponding plot for the THC kernel, $ \chi_{2} $ (in units of $ K_{B}T $) is double-peaked and numerically distinct from the $ \chi_{1} $ curve. For $ T = 300\, K $, observe how the double valleys of the $ \chi_{2} $ are now spread further and therefore contribute to the divergence of $ \mathcal{L}_{an} $ from $  \mathcal{L}_{0} $ at higher temperatures. This broadening between the valleys of the $ \chi_{2} $ curve is of the order of magnitude of $ k_{B}T $. However, for $ T = \lbrace 1\, K, 10\, K\rbrace $, (the low temperature range), the profiles for both AHC and THC kernels almost coincide (the pointed spikes around the $ E - \mu = 0 $ mark) imparting sufficient closeness between $ \mathcal{L}_{an} $ and $ \mathcal{L}_{0} $. Note that in the original formulation of the WFL, the ratio $ \mathcal{L}_{0} =  2.44 \times 10^{-8} W\Omega K^{-2} $ is exact only at $ T = 0 $ as the two kernels, $ \chi_{1} $ and $ \chi_{2} $ completely overlap. The deviation of $ \mathcal{L}_{an} $ vis-\'a-vis $\mathcal{L}_{0} $ is also enhanced by the non-linear dependence of the Berry curvature $\left(\Omega\right)$ on energy (Fig.~\ref{dberry}). The ratio of the integral of the products of $ \Omega\left(k\right)$ with the appropriate kernel $\left(\chi_{1} \text{and} \chi_{2}\right)$ in a window spanning several $ k_{B}T $, from which we determine the Lorenz number, is therefore strongly dependent on the energy and thus further reinforces the high temperature divergence from $ \mathcal{L}_{0} $, as shown in Fig.~\ref{lan}. The divergence, however, will be reduced in systems where the topologically-governed $ \Omega $ linearly increases with energy in the range of chosen $ k_{B}T $ values. The $ \Omega\left(k\right)$ evidently plays the role of transport parameters linked to scattering events~\cite{mahan2013many} used in the quantitative estimation of the classical Lorenz number.

Lastly, observe that the curves in Fig.~\ref{lan} clearly point to the role of anisotropy $\left(\nu\right)$, the band gap $\left(\Delta\right)$, and $ \mu $ - an outcome which is easily reconciled by remembering that $ \nu $ and $ \Delta $ adjust the dispersion profile (Eq.~\ref{diseq}). The $ \mu $ acts to alter the Fermi distribution (FD) by rearranging the occupation of the energy states contained in the dispersion curve. The FD enters the analysis through the previously defined kernel functions. Lastly, note that while we do not explicitly point out in Fig.~\ref{lan}, the inherent tilt $\left(\hbar v_{t}\sigma_{0}k_{y}\right)$ in the in $\beta $-borophene Hamiltonian (Eq.~\ref{boeqn}) serves as an ancillary tool to adjust $ \mathcal{L}_{an} $. A set of distinct tilts can suitably modify the attendant dispersion - similar to $\nu $ and $ \Delta $ - yielding a different $ \mathcal{L}_{an} $ in each case. 

In closing, we established the anomalous Lorenz number for massive and tilted Dirac systems using $ \beta $-borophene as a representative material. The temperature dependence of $ \mathcal{L}_{an} $, in contrast to the WFL-predicted constant $ \mathcal{L}_{0} $ was explained as the conjoined influence of kinetic transport parameters from the Boltzmann formalism and the topologically-induced Berry curvature. The tilt and anisotropy of the Dirac bands were also found relevant to adjust $ \mathcal{L}_{an} $ quantitatively. These results hold promise as topological materials - like borophene - with engineered $ \mathcal{L}_{an} $ can effectively tune the flow of transverse thermal currents as have been recently attempted with rare-earth magnets and thin-film devices.~\cite{li2017anomalous,das2019systematic} Lastly, values of $ \mathcal{L}_{an} $ lower than that of the classical Lorenz number indicate the possibility of an enhanced transverse electric conductivity or a reduced strength of the thermal counterpart - both essential design attributes in optimizing performance of nanoscale thermal devices.


\end{document}